\documentclass[prb,twocolumn]{revtex4}
\usepackage{graphicx} 

\usepackage{amsmath,amsthm,amssymb,mathrsfs}
\usepackage{bm}

\begin{document}

\title{Switching induced by spin Hall effect in an in-plane magnetized ferromagnet with the easy axis parallel to the current}

\author{Tomohiro Taniguchi
      }
 \affiliation{
 National Institute of Advanced Industrial Science and Technology (AIST), Spintronics Research Center, Tsukuba, Ibaraki 305-8568, Japan
 }

\date{\today} 
\begin{abstract}
{
Magnetization switching in a fine-structured ferromagnet of nanoscale by the spin-transfer torque excited via the spin Hall effect has attracted much attention 
because it enables us to manipulate the magnetization without directly applying current to the ferromagnet. 
However, 
the switching mechanism is still unclear in regard to the ferromagnet having an in-plane easy axis parallel to the current. 
Here, we develop an analytical theory of the magnetization switching in this type of ferromagnet, and reveal the threshold current formulas for a deterministic switching. 
It is clarified that the current should be in between a certain range determined by two threshold currents 
because the spin-transfer torque due to a large current outside the range brings the magnetization in an energetically unstable state, and causes magnetization precession around the hard axis. 
}
\end{abstract}

 \maketitle


\section{Introduction}
\label{sec:Introduction}

Spin-orbit interaction in a nonmagnetic metal scatters conducting electrons in a direction perpendicular to both current and spin angular momentum. 
Spin Hall effect is a physical phenomenon generating pure spin current via such spin-orbit interaction \cite{dyakonov71,hirsch99,kato04}. 
Attaching a ferromagnet to the nonmagnet, the pure spin current injected into the ferromagnet excites spin-transfer torque \cite{slonczewski96,berger96} 
and induces magnetization dynamics such as switching and auto-oscillation \cite{liu12a,liu12b,liu12c,pai12,cubukcu14,yu14,you15,torrejon15,fukami16a,oh16,brink16,lau16,fukami16b}. 
These magnetization dynamics have gained much attention from viewpoints of both fundamental and applied physics 
because these dynamics are interesting examples of nonlinear phenomena in nanoscale \cite{bertotti09}, and can be used for practical devices such as three-terminal magnetic random access memory \cite{dieny16}. 


The physical systems related to the spin Hall effect are classified into three different types. 
The first one is the system with a ferromagnet having an in-plane easy axis  orthogonal to the current direction \cite{liu12a,liu12c,pai12} named as type Y in Ref. \cite{fukami16b}. 
In this case, the spin polarization of the spin current is parallel to the easy axis. 
The theoretical analysis for the system is well developed 
because the system is similar to two-terminal magnetic multilayers proposed for the original concept of the spin-transfer torque \cite{slonczewski96,berger96,katine00,kiselev03,sun00,grollier03,kubota05}, 
where the magnetization dynamics is excited as a result of the competition between the spin-transfer torque and the damping torque. 
The second one is the system with a ferromagnet having a perpendicular easy axis \cite{liu12b,liu12c,cubukcu14,yu14,you15,torrejon15,fukami16a,oh16,brink16,lau16,kim12,kim16}, named as type Z \cite{fukami16b}. 
The magnetization switching in this system has been extensively studied \cite{lee13,lee14,taniguchi15PRB,taniguchi15,taniguchi19,lee20,zhu20}, particularly due to its applicability to practical devices. 
The third one is the system with a ferromagnet having the easy axis parallel to the current direction, named as type X \cite{fukami16b}. 
This system is easy to fabricate, and a small cross-section area for the current injection is also preferable for practical purposes. 
In addition, it was shown that the system is able to achieve a fast magnetization switching \cite{fukami16b}. 
Despite these fascinating properties, however, the physical mechanism of the magnetization switching in this system has not been investigated yet, 
which prevents us from establishing a comprehensive dynamical theory of the magnetization. 


The purpose of this work is to develop theoretical formulas of the magnetization switching caused by the spin Hall effect in the ferromagnet having the easy axis parallel to the current. 
Solving the Landau-Lifshitz-Gilbert (LLG) equation numerically, phase diagrams clarifying the magnetization state in a steady state are obtained. 
An analytical formula is derived for the critical current inducing the magnetization switching, showing good agreement with the phase diagram. 
The magnetization state after turning off the current is also investigated. 
The numerical simulation reveals the existence of another threshold value of the current for the switching, 
i.e., a deterministic switching can be achieved when the current is in between a range determined by two threshold currents. 
An analytical formula for this second threshold is also obtained. 
The results contribute to building a comprehensive view of the magnetization manipulation by the spin Hall effect. 


The paper is organized as follows. 
In Sec. \ref{sec:System description} we provide a system description, and show the phase diagram of the magnetization in a steady state. 
In Sec. \ref{sec:Critical current for instability threshold} the critical current formula destabilizing the magnetization from the initial state is derived. 
In Sec. \ref{sec:Deterministic switching} the condition for a deterministic switching after turning off the current is studied. 
In Sec. \ref{sec:Comparison to other systems} the comparison of the critical current with that in the different systems are discussed. 
Section \ref{sec:Conclusion} summarizes the conclusion of this work. 



\begin{figure}
\centerline{\includegraphics[width=1.0\columnwidth]{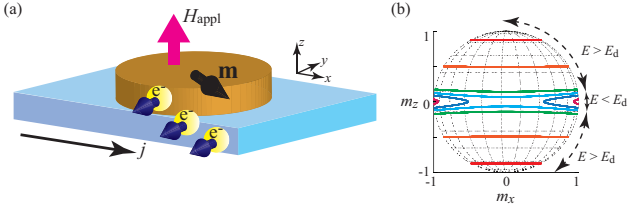}}
\caption{
         (a) Schematic view of the system. 
             The unit vector pointing in the magnetization direction is $\mathbf{m}$. 
             The electric current density and the external field are denoted as $j$ and $H_{\rm appl}$, respectively. 
         (b) Schematic view of the constant energy curve in the absence of the external field. 
         \vspace{-3ex}}
\label{fig:fig1}
\end{figure}



\section{System description}
\label{sec:System description}

The system we consider is schematically shown in Fig. \ref{fig:fig1}(a). 
The unit vector pointing in the magnetization direction of the ferromagnet is denoted as $\mathbf{m}$. 
The $z$ axis is perpendicular to the plane, while the $x$ axis is parallel to the electric current density $j$ in the nonmagnet. 
The spin Hall effect in the bottom nonmagnet generates pure spin-current flowing in the $z$ direction with the spin polarization in the $y$ direction, 
and excites the spin-transfer torque. 
An external field $H_{\rm appl}$ is applied in the $z$ direction. 
The magnetization dynamics in the ferromagnet is described by the LLG equation, 
\begin{equation}
  \frac{d \mathbf{m}}{dt}
  =
  -\gamma
  \mathbf{m}
  \times
  \mathbf{H}
  -
  \gamma
  H_{\rm s}
  \mathbf{m}
  \times
  \left(
    \mathbf{e}_{y}
    \times
    \mathbf{m}
  \right)
  +
  \alpha
  \mathbf{m}
  \times
  \frac{d \mathbf{m}}{dt},
  \label{eq:LLG}
\end{equation}
where $\gamma$ and $\alpha$ are the gyromagnetic ratio and the Gilbert damping constant, respectively. 
The magnetic field is given by 
\begin{equation}
  \mathbf{H}
  =
  H_{\rm K}
  m_{x}
  \mathbf{e}_{x}
  +
  \left(
    H_{\rm appl}
    -
    4\pi M 
    m_{z} 
  \right)
  \mathbf{e}_{z}, 
  \label{eq:field}
\end{equation}
where $H_{\rm K}$ is the in-plane magnetic anisotropy field in the $x$ direction, 
whereas $-4\pi M$ is the shape anisotropy field in the $z$ direction. 
The strength of the spin-transfer torque by the spin Hall effect is 
\begin{equation}
  H_{\rm s}
  =
  \frac{\hbar \vartheta j}{2eMd},
\end{equation}
where $\vartheta$ is the spin Hall angle of the nonmagnetic heavy metal, 
and $M$ and $d$ are the saturation magnetization and thickness of the free layer, respectively. 
The values of the parameters used in this work are derived from typical experiments and simulations of the spin Hall phenomena and/or magnetic multilayers as 
$M=1500$ emu/cm${}^{3}$, $H_{\rm K}=200$ Oe, $\vartheta=0.4$, $d=1.0$ nm, $\gamma=1.764 \times 10^{7}$ rad/(Oe s), and $\alpha=0.005$ 
\cite{torrejon15,taniguchi15,taniguchi19}; see also Appendix \ref{sec:AppendixA} for the calculation details. 


We note that the macrospin model is used to derive the analytical formulas of the threshold currents. 
This model has been used to analyze the switching dynamics in both two-terminal and three-terminal devices \cite{sun00,grollier03,lee14,taniguchi15,taniguchi19,lee20}. 
It is known that macrospin model is applicable to small devices. 
For example, the applicability of the model to the two-terminal perpendicularly magnetized system was investigated in Refs. \cite{sun11,oflynn13,oflynn15}. 
The applicability of the macrospin model for the present three-terminal was verified in the experiment carried out in Refs. \cite{fukami16b}, 
where the thickness of the free layer was 1.48 nm whereas the short and long axes, corresponding to the $y$ and $x$ directions in the present geometry, were 160 and 400 nm, respectively. 
Therefore, we believe that using the macrospin model is reasonable to analyze the switching behavior of the present system. 


For the latter discussion, it is useful to introduce the magnetic energy density $E$ as 
$E=-M \int d \mathbf{m}\cdot\mathbf{H}$, 
\begin{equation}
  E
  =
  -MH_{\rm appl}
  m_{z}
  -
  \frac{MH_{\rm K}}{2}
  m_{x}^{2}
  +
  2\pi M^{2}
  m_{z}^{2}.
  \label{eq:energy}
\end{equation}
Note that the energy density $E$ has two minima corresponding to 
\begin{equation}
  \mathbf{m}_{0\pm}
  =
  \begin{pmatrix}
    \pm m_{0x} \\
    0 \\
    m_{0z}
  \end{pmatrix},
  \label{eq:initial_state}
\end{equation}
where $m_{0z}=H_{\rm appl}/(H_{\rm K}+4\pi M)$ and $m_{0x}=\sqrt{1-m_{0z}^{2}}$. 
The minimum energy density is 
\begin{equation}
  E_{\rm min}
  =
  -M
  \frac{H_{\rm appl}^{2}+H_{\rm K}(H_{\rm K}+4\pi M)}{2(H_{\rm K}+4\pi M)}.
  \label{eq:energy_minimum}
\end{equation}
The saddle point of the energy density locates at 
\begin{equation}
  \mathbf{m}_{{\rm d}\pm}
  =
  \begin{pmatrix}
    0 \\
    \pm m_{{\rm d}y} \\
    m_{{\rm d}z}
  \end{pmatrix},
\end{equation}
where $m_{{\rm d}z}=H_{\rm appl}/(4\pi M)$ and $m_{{\rm d}y}=\sqrt{1-m_{{\rm d}z}^{2}}$. 
The saddle-point energy density is 
\begin{equation}
  E_{\rm d}
  =
  -\frac{MH_{\rm appl}^{2}}{8\pi M}.
  \label{eq:energy_saddle}
\end{equation}
An example of the constant energy line is shown in Fig. \ref{fig:fig1}(b), where the external field is assumed to be zero; 
see also Appendix \ref{sec:AppendixA} for the values of the energy density. 
In this case ($H_{\rm appl}=0$) in particular, the minimum points locate at $\mathbf{m}_{0\pm}=\pm\mathbf{e}_{x}$, 
whereas the saddle points are $\mathbf{m}_{{\rm d}\pm}=\pm\mathbf{e}_{y}$. 
These minimum and saddle points play significant role for the determination of the switching condition, as discussed below. 



\begin{figure}
\centerline{\includegraphics[width=1.0\columnwidth]{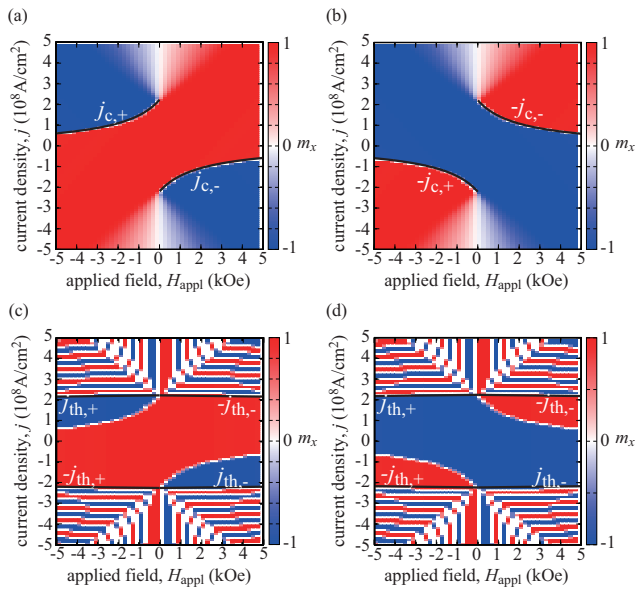}}
\caption{
         Phase diagrams of $m_{x}$ in a steady state (a),(b) in the presence of the current, and (c),(d) after turning off the current. 
         The initial states are $\mathbf{m}_{0+}$ in (a) and (c), whereas it is $\mathbf{m}_{0-}$ in (b) and (d). 
         The black lines with labels $\pm j_{\rm c,\pm}$ and $\pm j_{\rm th,\pm}$ in (a),(b) and (c),(d) correspond to Eqs. (\ref{eq:jc_finite_field}) and (\ref{eq:jth}), respectively. 
         \vspace{-3ex}}
\label{fig:fig2}
\end{figure}



Figure \ref{fig:fig2}(a) summarizes $m_{x}$ in a steady state in the presence of the current. 
The initial state is $\mathbf{m}_{0+}$, which is close to $m_{x}=1$ shown by the red color. 
The magnetization stays near the initial state around small current and/or field regions. 
On the other hand, when the current magnitude becomes larger than critical values, the magnetization moves from the initial state. 
The critical current density is positive for $H_{\rm appl}<0$, whereas it becomes negative for $H_{\rm appl}>0$. 
In this paper, we denote these critical current densities as $j_{{\rm c},+}$ and $j_{{\rm c},-}$, as shown in Fig. \ref{fig:fig2}(a). 
Figure \ref{fig:fig2}(b) shows $m_{x}$ in a steady state, where the initial state is $\mathbf{m}_{0-}$ close to $m_{x}=-1$ shown by the blue color. 
Similar to Fig. \ref{fig:fig2}(a), the magnetization moves from the initial state when the current density exceeds the critical value. 
In this case, the critical current is positive for $H_{\rm appl}>0$, whereas it becomes negative for $H_{\rm appl}<0$. 
The first purpose of the following discussion is to derive a theoretical formula of the critical currents to move the magnetization from the initial state. 

It should also be noted that the magnetization state after turning off the current is also of interest, in particular for practical applications. 
Note that the magnetization finally saturates to $\mathbf{m}_{0+}$ or $\mathbf{m}_{0-}$ because these are energetically stable states. 
Figure \ref{fig:fig2}(c) shows $m_{x}$ in a steady state after the current is turned off, where the initial state is $\mathbf{m}_{0+}$. 
Comparing Fig. \ref{fig:fig2}(c) with Fig. \ref{fig:fig2}(a), 
the existence of another threshold current for the switching is found. 
For example, the relaxed state of the magnetization in Fig. \ref{fig:fig2}(c) is $\mathbf{m}_{0-}$ ($m_{x}\simeq -1$ shown by the blue color) 
when the current density is slightly larger than the critical current density determining the boundary between the states of $m_{x}\simeq 1$ and $m_{x}\simeq -1$ found in Fig. \ref{fig:fig2}(a). 
However, when the magnitude of the current density further increases, both $\mathbf{m}_{0+}$ (red) and $\mathbf{m}_{0-}$ (blue) states coexist, as can be seen in a large current region of Fig. \ref{fig:fig2}(c). 
This fact indicates that, even if the magnetization direction moves to a position close to the switching state, which is $m_{x}\simeq -1$ in this case, by the spin-transfer torque, 
it moves back to the initial state ($m_{x}\simeq -1$), depending on the value of the current density. 
Therefore, the current density should be smaller than a certain value for a deterministic switching. 
In this paper, we denote the threshold current density determining the upper boundary of the deterministic switching as $j_{{\rm th},\pm}$, as can be seen in Fig. \ref{fig:fig2}(c). 
A similar behavior is observed when the initial state is $\mathbf{m}_{0-}$, as shown in Fig. \ref{fig:fig2}(d). 
The second purpose of the following discussion is to clarify the origin of these complex phase diagram and to derive the threshold current formula for the deterministic switching. 



\begin{figure*}
\centerline{\includegraphics[width=2.0\columnwidth]{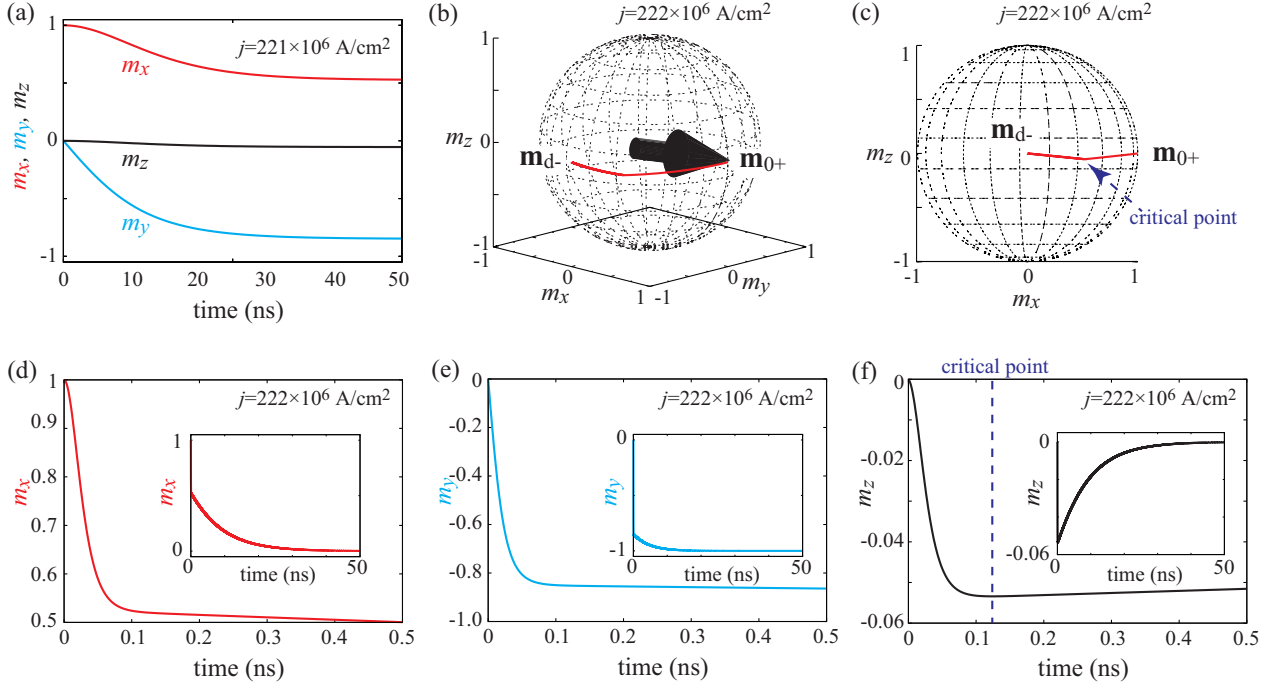}}
\caption{
         (a) Time evolutions of the magnetization components for $j=221 \times 10^{6}$ A/cm${}^{2}$, which is slightly below the critical value. 
         (b) and (c) Dynamic trajectories from different viewpoints, where the current density $j=222\times 10^{6}$ A/cm${}^{2}$ is larger than the critical value. 
         (d)-(f) Time evolutions of the magnetization components (red for $m_{x}$, blue for $m_{y}$, and black for $m_{z}$) for $j=222\times 10^{6}$ A/cm${}^{2}$. 
         The insets show the time evolutions in a long time range, indicating that the magnetization finally saturates to $\mathbf{m}=-\mathbf{e}_{y}$. 
         The dotted arrow in (c) and line in (f) indicate the position satisfying $dm_{z}/dt=0$, which is called a critical point. 
         The external magnetic field is zero in these figures. 
         \vspace{-3ex}}
\label{fig:fig3}
\end{figure*}



\section{Critical current for instability threshold}
\label{sec:Critical current for instability threshold}

In this section, we derive the theoretical formula of the critical current density to move the magnetization from the initial state. 
First, we study the case of the zero-field switching. 
Second, the theory is extended to the case of the finite-field switching. 
The role of the thermal fluctuation at finite temperature is also discussed. 


\subsection{Switching condition for zero-field case}
\label{sec:Switching condition for zero-field case}

Let us first investigate the magnetization dynamics in the absence of the external magnetic field. 
In the meantime we consider the dynamics caused by a positive current, and assume that the initial state is $\mathbf{m}_{0+}$, 
which is parallel to the $x$ axis as $\mathbf{m}_{0+}=+\mathbf{e}_{x}$ for $H_{\rm appl}=0$. 


We should first note the definition of the critical current density for the zero-field case. 
The spin-transfer torque due to the spin Hall effect moves the magnetization parallel to the $y$ axis. 
The precession torque due to the magnetic field also becomes zero when the magnetization is parallel to the $y$ axis 
because the point corresponds to the saddle point of the energy density. 
Therefore, the critical current density in the zero-field case is defined as a current density over which the magnetization saturates to the saddle point. 


Figure \ref{fig:fig3}(a) shows the time evolutions of the magnetization components, $m_{x}$ (red), $m_{y}$ (blue), and $m_{z}$ (black), 
when the current density $j=221\times 10^{6}$ A/cm${}^{2}$ is slightly smaller than the critical value. 
Although the magnetization moves from the initial state, it does not reach the saddle point. 
In particular, $m_{z}$ moves from the initial state ($m_{z}=0$) to a certain value monotonically. 


On the other hand, Fig. \ref{fig:fig3}(b) shows the dynamic trajectory of the magnetization when the current density $j=222 \times 10^{6}$ A/cm${}^{2}$ is larger than the critical value. 
Starting from the initial state $\mathbf{m}_{0+}$, the magnetization saturates to the saddle point $\mathbf{m}_{{\rm d}-}$. 
Figure \ref{fig:fig3}(c) shows the dynamic trajectory observed from a different view angle. 
The figure indicates that the magnetization first moves to the negative $z$ direction, and then turns back to the $xy$ plane 
because $m_{z}$ is zero for both the initial ($\mathbf{m}_{0+}$) and final ($\mathbf{m}_{{\rm d}-}$) states. 
This fact can also be confirmed from Figs. \ref{fig:fig3}(d)-\ref{fig:fig3}(f), where the time evolutions of the magnetization components are shown. 
Whereas $m_{x}$ and $m_{y}$ shown in Figs. \ref{fig:fig3}(d) and \ref{fig:fig3}(e) monotonically moves from the initial to the final state, 
$m_{z}$ in Fig. \ref{fig:fig3}(f) shows a local minimum at which $dm_{z}/dt=0$. 
The inset in Figs. \ref{fig:fig3}(d)-\ref{fig:fig3}(f) indicate that the magnetization finally saturates to the saddle point, $\mathbf{m}_{{\rm d}-}=-\mathbf{e}_{y}$. 
Comparing Figs. \ref{fig:fig3}(c)-\ref{fig:fig3}(f) with Fig. \ref{fig:fig3}(a), 
we conclude that the existence of the point satisfying $dm_{z}/dt=0$ determines a bifurcation between the switching and non-switching states. 
Therefore, let us call the point of $dm_{z}/dt=0$ as a critical point, for convention; 
see also Figs. \ref{fig:fig3}(c) and \ref{fig:fig3}(f), where the positions of the critical point 
in the real and time-domain spaces are shown, respectively. 


Figure \ref{fig:fig3}(b) also indicates that the switching occurs without accompanying magnetization precession. 
This is in contrast to the original idea of the spin-transfer torque switching \cite{slonczewski96,sun00,grollier03}, 
where the spin-transfer torque compensates for the damping torque, and therefore, the precession torque due to the magnetic field, 
corresponding to the first term on the right-hand side of Eq. (\ref{eq:LLG}), is dominant.  
The fact that the switching in the present system does not accompany precession dynamics implies that the switching occurs as a result of 
the competition between the precession torque and the spin-transfer torque. 
Therefore, the damping torque, which is proportional to the small constant $\alpha$, can be neglected to determine the switching condition. 
In fact, we confirmed from the numerical simulation that the critical current is unchanged even when we use a damping value of $\alpha=0.02$, 
which is four times larger than that used in Fig. \ref{fig:fig3}. 


We note that the monotonic evolutions of $m_{x}$ and $m_{y}$ shown in Figs. \ref{fig:fig3}(d) and \ref{fig:fig3}(e) indicate that 
the signs of $dm_{x}/dt$ and $dm_{y}/dt$ are fixed. 
Therefore, the following conditions should be satisfied, 
\begin{equation}
  \frac{1}{\gamma}
  \frac{dm_{x}}{dt}
  \simeq 
  \left(
    4\pi M 
    m_{z}
    +
    H_{\rm s}
    m_{x}
  \right)
  m_{y}
  <
  0,
  \label{eq:zero_field_condition_1}
\end{equation}
\begin{equation}
  \frac{1}{\gamma}
  \frac{dm_{y}}{dt}
  \simeq
  -\left(
    H_{\rm K}
    +
    4\pi M
  \right)
  m_{x}
  m_{z}
  -
  H_{\rm s}
  \left(
    m_{x}^{2}
    +
    m_{z}^{2}
  \right)
  <
  0,
  \label{eq:zero_field_condition_2}
\end{equation}
Equations (\ref{eq:zero_field_condition_1}) and (\ref{eq:zero_field_condition_2}) should be satisfied for the switching process of 
$0 \le m_{x} \lesssim 1$ and $-1 \lesssim m_{y} \le 0$. 
Note that the critical point satisfying $dm_{z}/dt=0$ exists when the switching occurs, as mentioned above. 
The LLG equation for $m_{z}$ indicates that the critical point is determined by the condition, 
\begin{equation}
  H_{\rm K}
  m_{x}
  +
  H_{\rm s}
  m_{z}
  =
  0. 
  \label{eq:critical_point}
\end{equation}
Substituting Eq. (\ref{eq:critical_point}) into Eqs. (\ref{eq:zero_field_condition_1}) and (\ref{eq:zero_field_condition_2}), 
we find that the switching condition at the critical point becomes 
\begin{equation}
  \left(
    H_{\rm s}
    -
    \frac{H_{\rm K}4\pi M}{H_{\rm s}}
  \right)
  m_{x}
  >
  0.
  \label{eq:zero_field_switching_condition}
\end{equation}
Since $m_{x}>0$ during the switching in the present case, 
we find that Eq. (\ref{eq:zero_field_switching_condition}) becomes $H_{\rm s}>\sqrt{H_{\rm K}4\pi M}$, 
which gives the critical current formulas as 
\begin{equation}
  j_{\rm c}
  =
  \frac{2 e Md}{\hbar \vartheta}
  \sqrt{
    H_{\rm K}
    4\pi M
  }.
  \label{eq:jc_zero_field}
\end{equation}
The value of Eq. (\ref{eq:jc_zero_field}) for the present calculation is $221 \times 10^{6}$ A/cm${}^{2}$, 
which shows good agreement with the numerical simulation shown in Fig. \ref{fig:fig3}. 
Considering that the above discussion focuses on the switching by positive current, 
it should be mentioned that negative current can also induce the switching, where the critical current density is given by $-j_{\rm c}$. 
We denote Eq. (\ref{eq:jc_zero_field}) as $j_{\rm c,+}$, whereas $-j_{\rm c}$ is denoted as $j_{\rm c,-}$. 
When the initial state is $\mathbf{m}_{0-}$, $-j_{\rm c,\pm}$ are the critical current densities. 


We note that Eq. (\ref{eq:jc_zero_field}) can be derived from Eq. (\ref{eq:zero_field_switching_condition}) 
without determining the value of $m_{x}$ at the critical point. 
It is, however, useful to evaluate its value for the latter discussion. 
We find that $m_{x}$ at the critical point is well approximated as (see also Appendix \ref{sec:AppendixB}) 
\begin{equation}
  m_{x}
  =
  \frac{1}{2}. 
  \label{eq:mx_critical_point_zero_field}
\end{equation}
The numerically evaluated value of $m_{x}$ at the critical point is $0.520$ [see Fig. \ref{fig:fig3}(d) and Appendix \ref{sec:AppendixB}], 
which is close to Eq. (\ref{eq:mx_critical_point_zero_field}). 
In the next section, we extend Eq. (\ref{eq:mx_critical_point_zero_field}) to the finite field case to derive the critical current formula. 


At the end of this section, we would like to provide a brief comment on two timescales observed in Fig. \ref{fig:fig3}. 
Whereas the magnetization moves from the initial state to the critical point fast, 
it takes a long time to saturate from the critical point to the fixed point ($\mathbf{m}\to -\mathbf{e}_{y}$). 
This result is in contrast with the spin-transfer torque switching in two-terminal devices, as well as that in the three-terminal type Y-device, 
where the switching time has a simple relation to the current as $1/t \propto j-j_{\rm c}$ \cite{sun00,suzuki09}. 
In addition, long-time dynamics as such are not observed for the switching in the presence of external magnetic field, as shown in next section. 
We note that investigating the origin of two timescales is of interest, and further work will be necessary in future. 




\begin{figure}
\centerline{\includegraphics[width=1.0\columnwidth]{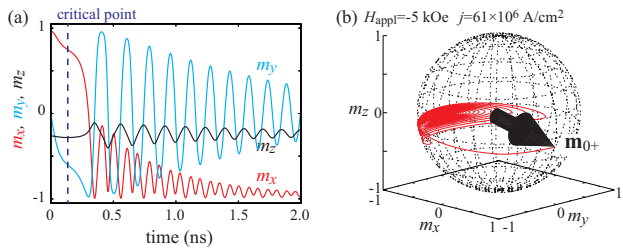}}
\caption{
         (a) Time evolutions of the magnetization components and (b) its trajectory from $\mathbf{m}_{0+}$ for $j=61 \times 10^{6}$ A/cm${}^{2}$ and $H_{\rm appl}=-5$ kOe. 
             The dotted line in (a) indicates the position of the critical point satisfying $dm_{z}/dt=0$. 
         \vspace{-3ex}}
\label{fig:fig4}
\end{figure}



\subsection{Switching condition for finite field case}
\label{sec:Switching condition for finite field case}

Let us extend Eq. (\ref{eq:jc_zero_field}) to the system in the presence of a finite external field. 
Figures \ref{fig:fig4}(a) and \ref{fig:fig4}(b) show the time evolutions of the magnetization components and the dynamic trajectory 
for $j=61 \times 10^{6}$ A/cm${}^{2}$ and $H_{\rm appl}=-5.0$ kOe. 
The initial state is $\mathbf{m}_{0+}$. 
In comparison with the zero-field case shown in Fig. \ref{fig:fig3}, 
the magnetization can move close to the switched state ($\mathbf{m}_{0-}\simeq -\mathbf{e}_{x}$). 
This is because the points at which the spin-transfer torque and the precession torque due to the magnetic field become zero are different from the case without external magnetic field. 
We also notice that the magnetization does not show a precessional motion before ($t \lesssim 0.3$ ns) reaching the boundary ($m_{x}=0$) between two stable states, 
although the precessional motion appears after the magnetization overcomes the boundary. 
This fact indicates that the critical current is mainly determined by the competition between the precession and spin-transfer torques, 
as in the case of the zero-field switching. 
In fact, we confirm that the critical current density is $63 \times 10^{6}$A/cm${}^{2}$ for $\alpha=0.02$, which is close to that for $\alpha=0.005$, 
indicating that the damping torque does not play a critical role to determine the instability threshold. 
The fact that the switching occurs without accompanying the precessional motion also indicates that 
the conditions that $dm_{x}/dt$ and $dm_{y}/dt$ have fixed signs, used in the derivation of Eq. (\ref{eq:jc_zero_field}) 
can also be used for the derivation of the critical current formula. 
Then, we find that the critical current density can be determined by the following condition; 
\begin{equation}
  -H_{\rm appl}
  +
  \left(
    H_{\rm s}
    -
    \frac{H_{\rm K}4\pi M}{H_{\rm s}}
  \right)
  m_{x}
  >
  0.
  \label{eq:finite_field_switching_condition}
\end{equation}
Equation (\ref{eq:finite_field_switching_condition}) becomes Eq. (\ref{eq:zero_field_switching_condition}) in the limit of $H_{\rm appl}\to 0$. 
We should, however, evaluate the value of $m_{x}$ at the critical point for the finite field case, 
in contrast with the derivation of Eq. (\ref{eq:jc_zero_field}) where evaluating the value of $m_{x}$ in Eq. (\ref{eq:zero_field_switching_condition}) was not necessary. 
For example, the critical point satisfying $dm_{z}/dt=0$ appears at $t=0.133$ ns for $j=61\times 10^{6}$A/cm${}^{2}$ and $H_{\rm appl}=-5.0$kOe, as shown by the dotted line in Fig. \ref{fig:fig4}(a). 
To proceed with the analysis, we extend the derivation of Eq. (\ref{eq:mx_critical_point_zero_field}) to the finite field case, 
and find that $m_{x}$ at the critical point is given by (see also Appendix \ref{sec:AppendixB}) 
\begin{widetext}
\begin{equation}
  m_{x}
  =
  \frac{H_{\rm K}(H_{\rm K}+4\pi M)H_{\rm s}}{2H_{\rm K}H_{\rm s}\sqrt{(H_{\rm K}+4\pi M+H_{\rm appl})(H_{\rm K}+4\pi M-H_{\rm appl})} + H_{\rm appl}(H_{\rm s}^{2}-H_{\rm K}^{2})}.
  \label{eq:mx_critical_point_finite_field}
\end{equation}
\end{widetext}
For example, the value of Eq. (\ref{eq:mx_critical_point_finite_field}) for $j=61\times 10^{6}$ A/cm${}^{2}$ and $H_{\rm appl}=-5.0$ kOe is 0.755, 
which is close to the result of the numerical simulation in Fig. \ref{fig:fig3}(a), where $m_{x}=0.750$ at the critical point appeared. 
Substituting Eq. (\ref{eq:mx_critical_point_finite_field}) into Eq. (\ref{eq:finite_field_switching_condition}), 
we find that the critical current density for the finite field case is given by 
\begin{widetext}
\begin{equation}
  j_{\rm c,\pm}
  =
  \pm
  \frac{2eMd}{\hbar\vartheta}
  \frac{H_{\rm K}(H_{\rm K}+4\pi M)\sqrt{H_{\rm K}4\pi M} \pm H_{\rm appl}H_{\rm K} \sqrt{(H_{\rm K}+4\pi M+H_{\rm appl})(H_{\rm K}+4\pi M-H_{\rm appl})}}
    {H_{\rm K}(H_{\rm K}+4\pi M)-H_{\rm appl}^{2}}
  \label{eq:jc_finite_field}
\end{equation}
\end{widetext}
Note that $j_{\rm c.+}$ is the critical current density for $H_{\rm appl}>0$, 
whereas $j_{\rm c,-}$ is that for $H_{\rm appl}<0$. 
On the other hand, when the initial state is $\mathbf{m}_{0-}$,
the critical current density for $H_{\rm appl}>(<)0$ is $-j_{\rm c,-(+)}$. 
The value of $j_{\rm c,+}$ estimated from Eq. (\ref{eq:jc_finite_field}) is $59 \times 10^{6}$A/cm${}^{2}$ for $H_{\rm appl}=-5.0$kOe, 
which shows good agreement with the result of the numerical simulation where the switching occurs for the current density of $j=61 \times 10^{6}$A/cm${}^{2}$, as shown in Fig. \ref{fig:fig4}(a). 




\begin{figure*}
\centerline{\includegraphics[width=2.0\columnwidth]{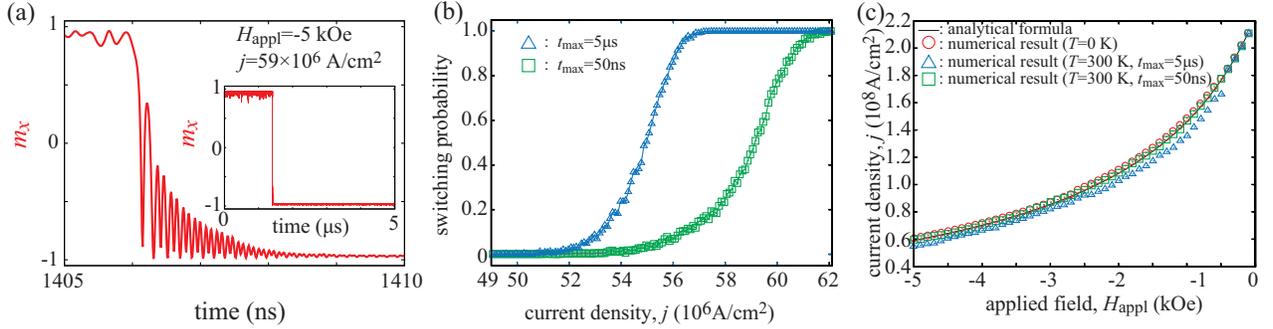}}
\caption{
         (a) An example of the magnetization dynamics at finite temperature, where $H_{\rm appl}=-5.0$ kOe and $j=59\times 10^{6}$ A/cm${}^{2}$. 
             The main figure shows the time evolution of $m_{x}$ near the switching, whereas the inset shows that over a long range of time. 
         (b) Dependences of the switching probabilities $P(j)$ on the current density for the current pulses of $5$ $\mu$s (blue triangles) and $50$ ns (green squares). 
             The switching probability is evaluated from the 1000 trials of the LLG equation, where the initial state is reset at every trial.
         (c) The dependencies of the current density maximizing $dP(j)/dj$ on the external magnetic field 
             for the long (blue triangles) and short (green squares) pulses. 
             For comparison, the critical current densities at zero temperature, estimated from Eq. (\ref{eq:jc_finite_field}) (black line) and the numerical simulation (red circles) 
             in Sec. \ref{sec:Switching condition for finite field case} are also shown. 
         \vspace{-3ex}}
\label{fig:fig5}
\end{figure*}



\subsection{Dynamics at finite temperature}
\label{sec:Dynamics at finite temperature}

This work mainly focuses on the magnetization dynamics at zero temperature to derive the critical current formula for the switching. 
Simultaneously, however, the role of the thermal fluctuation at finite temperature \cite{brown63} is also of great interest. 
In this section, we discuss the magnetization dynamics at finite temperature by solving the LLG equation with the random torque numerically. 
In particular, we focus on the current range near the critical value $j_{\rm c}$ 
where the thermal fluctuation is expected to play a critical role on the magnetization switching 
\cite{myers02,koch04,apalkov05,suzuki09,bedau10,taniguchi11,butler12,taniguchi12,sato12,taniguchi12APEX,newhall13,tomita13,taniguchi13,gopman14,lee14}. 

The thermal fluctuation provides a torque, 
\begin{equation}
  \bm{\tau}
  =
  -\gamma
  \mathbf{m}
  \times
  \mathbf{h},
  \label{eq:random_torque}
\end{equation}
which should be added to the right-hand side of Eq. (\ref{eq:LLG}). 
The components $h_{k}(t)$ ($i=x,y,z$) of the random field $\mathbf{h}$ obeys the fluctuation-dissipation theorem \cite{brown63}, 
\begin{equation}
  \langle 
    h_{k}(t)
    h_{\ell}(t^{\prime})
  \rangle 
  =
  \frac{2 \alpha k_{\rm B}T}{\gamma MV}
  \delta_{k\ell}
  \delta(t-t^{\prime}), 
  \label{eq:FDT}
\end{equation}
where $V=Sd$ is the volume of the free layer consisting of the cross section area $S$ and the thickness $d$. 
According to Ref. \cite{fukami16b} where the applicability of the macrospin model was confirmed, we assume that $S=\pi \times 80 \times 200$ nm${}^{2}$. 
The temperature $T$ is fixed to $T=300$ K. 
The calculation method to solve Eq. (\ref{eq:LLG}) with Eq. (\ref{eq:random_torque}) are summarized in Appendix \ref{sec:AppendixC}. 


Figure \ref{fig:fig5}(a) shows an example of the magnetization dynamics at finite temperature, 
where the external magnetic field and the current density are $H_{\rm appl}=-5.0$ kOe and $j=59 \times 10^{6}$ A/cm${}^{2}$, respectively. 
We remind the readers that the critical current density estimated from the numerical simulation at zero temperature is $61 \times 10^{6}$ A/cm${}^{2}$, 
as mentioned in Sec. \ref{sec:Switching condition for finite field case}.
The figure is an enlarged view near the switching, whereas the inset shows the time evolution of $m_{x}$ over a long range of the time ($0 \le t \le 5$ $\mu$s). 
Since the current density is smaller than the critical value at zero temperature, the magnetization stays near the initial state with random motion for a long time. 
When, however, the magnetization comes close to the critical point and the random torque further assists the motion to overcome the point, the switching is achieved. 


Next, let us show the dependence of the switching probability $P$ on the current density. 
We note that the existence of the random torque does not guarantee the switching, although Fig. \ref{fig:fig5}(a) shows an example of the probabilistic switching. 
We also emphasize that the time at which the switching occurs is random. 
In particular, we note that a long time is necessary to observe a finite probabilistic switching, especially for the current density much lower than the critical value at zero temperature. 
On the other hand, for practical purpose, a short current pulse is used to achieve a fast switching. 
Therefore, we calculate the switching probabilities for both long ($t_{\rm max}=5$ $\mu$s) and short ($t_{\rm max}=50$ ns) current pulses, 
where we regard the magnetization switched when, starting from the initial condition $m_{x}(t=0)>0$, the magnetization at $t=t_{\rm max}$ satisfies $m_{x}(t=t_{\rm max})<0$; 
see also Appendix \ref{sec:AppendixC} for the evaluation method of the switching probability. 
Figure \ref{fig:fig5}(b) shows the switching probabilities for the cases of the long (blue triangles) and short (green squares) pulses, where $H_{\rm appl}=-5.0$ kOe. 
For the case of the long pulse, the switching is observed even at the current density relatively smaller than the critical value ($61\times 10^{6}$A/cm${}^{2}$) evaluated by the LLG simulation at zero temperature. 
The switching probability shows a gradual evolution from $P=0$ to $P=1$. 
On the other hand, the switching probability increases near the critical current density for the short-pulse case. 
Figure \ref{fig:fig5}(c) summarizes the magnetic field dependence of the current density at which the switching probability increases most rapidly, 
i.e., the current density maximizing $dP(j)/dj$, which is regarded as the switching current density at finite temperature. 
For comparison, we also shows the critical current density at zero temperature, 
where the black line is obtained from Eq. (\ref{eq:jc_finite_field}) whereas the red circles are obtained by the numerical simulation, as done in Sec. \ref{sec:Switching condition for finite field case}. 
For the long-pulse ($t_{\rm max}=5$ $\mu$s) case, the switching current densities at finite temperature is reduced to roughly 90\% of those at zero temperature; 
for example, the switching current density for $H_{\rm appl}=-5.0$ kOe is estimated to be $54.8\times 10^{6}$ A/cm${}^{2}$, whereas that for zero-temperature case was $61\times 10^{6}$ A/cm${}^{2}$, as mentioned in Sec. \ref{sec:Switching condition for finite field case}. 
On the other hand, the switching current densities for the short-pulse ($t_{\rm max}=50$ ns) case are slightly lower but nearly overlap the zero-temperature results; 
for example, the switching current for $H_{\rm appl}=-5.0$ kOe is estimated to be $59.7\times 10^{6}$ A/cm${}^{2}$. 
These results indicate that, although the thermal fluctuation provides a possibility to switching the magnetization by the current lower than the critical value, 
the critical current formulas derived in this work will be useful to analyze the magnetization switching, 
particularly for a short-pulse switching required in practical applications. 


\subsection{Summary of this section}

In this section, we derive the critical current formula to move the magnetization from the initial state. 
The main results are Eqs. (\ref{eq:jc_zero_field}) and (\ref{eq:jc_finite_field}). 
Equation (\ref{eq:jc_zero_field}) is the critical current formula for the zero-field case, 
whereas Eq. (\ref{eq:jc_finite_field}) is the extension to the finite field case. 
Equation (\ref{eq:jc_finite_field}) reproduces Eq. (\ref{eq:jc_zero_field}) in the limit of $H_{\rm appl}\to 0$. 

The black solid lines in Figs. \ref{fig:fig2}(a) and \ref{fig:fig2}(b) are $j_{\rm c,\pm}$ and $-j_{\rm c,\pm}$. respectively. 
As shown, the formula of the critical current density given by Eq. (\ref{eq:jc_finite_field}) well explains 
the boundary between the initial and switching states in the presence of the current, 
showing the validity of Eq. (\ref{eq:jc_finite_field}). 




\begin{figure}
\centerline{\includegraphics[width=1.0\columnwidth]{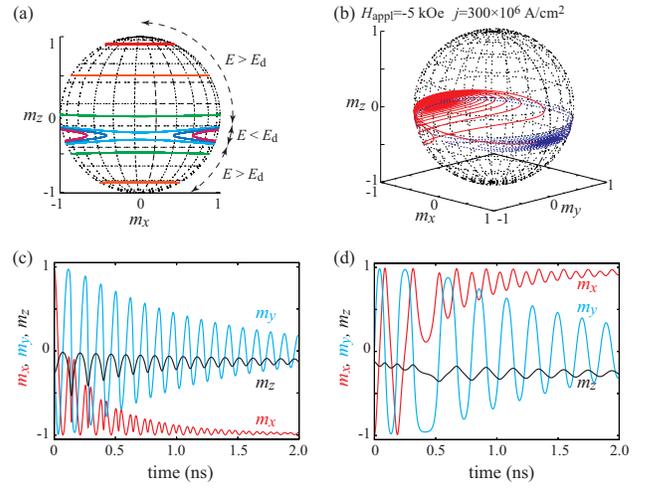}}
\caption{
         (a) Schematic view of the constant energy lines for $H_{\rm appl}=-5.0$ kOe. 
         (b) The red solid line is the trajectory of the magnetization for $j=300 \times 10^{6}$ A/cm${}^{2}$ and $H_{\rm appl}=-5$ kOe. 
             The blue dotted line represents the trajectory after turning off the current. 
             Time evolutions of the magnetization components in the presence of the current and after turning off the current are shown in (c) and (d), 
             where the origin of the time is set to be $t=0$ in (d), for simplicity. 
             Practically, the dynamics starts after turning off the current. 
         \vspace{-3ex}}
\label{fig:fig6}
\end{figure}



\section{Deterministic switching}
\label{sec:Deterministic switching}

Finally, we discuss the theoretical condition for the deterministic switching. 
In many cases including practical purposes, the magnetization state not only in the presence of the current but also the state after turning off the current is of great interest. 
The phase diagrams in Figs. \ref{fig:fig2}(c) and \ref{fig:fig2}(d) reveal the existence of the current range in which a deterministic switching is achieved, as mentioned in Sec. \ref{sec:System description}. 
For example, for $H_{\rm appl}=-5.0$ kOe, the deterministic switching occurs in the current range of $70 \le j \le 270$ $\times 10^{6}$A/cm${}^{2}$, as shown in Fig. \ref{fig:fig2}(c), 
where the lower boundary, $70\times 10^{6}$A/cm${}^{2}$ is well explained by Eq. (\ref{eq:jc_finite_field}). 
Here, we note that the current density in Fig. \ref{fig:fig2} is varies by $10\times 10^{6}$ A/cm${}^{2}$ step, and thus, 
the deterministic switching range here is mentioned as $70 \le j \le 270$ $\times 10^{6}$A/cm${}^{2}$. 
However, to be more precise as mentioned in Sec. \ref{sec:Critical current for instability threshold}.
the detail value of the lower boundary is $61 \times 10^{6}$ A/cm${}^{2}$. 
There is, in addition, another threshold value, which is $270 \times 10^{6}$A/cm${}^{2}$ in this example. 
Above this second threshold, both $\mathbf{m}_{0+}$ (red) and $\mathbf{m}_{0-}$ (blue) appear as a final state, as shown in Fig. \ref{fig:fig2}(c), 
which is sensitive to the values of the parameters such as the current and field. 
Let us derive the theoretical formula of the second threshold current here. 


Before the discussion, we briefly note that the word "deterministic" in this section is used to express 
the switching which is not sensitive to the values of the external magnetic field and the current density, 
and is not used to distinguish from the dynamics at finite temperature studied in Sec. \ref{sec:Dynamics at finite temperature}. 
Since the phase diagram in Fig. \ref{fig:fig2} is obtained by the zero-temperature simulation, the magnetization states are determined deterministically. 
The results in Fig. \ref{fig:fig2} reveal, however, that there are the regions where the relaxed state of the magnetization is sensitive to the values of the external magnetic field and the current density, even at zero temperature. 
Such a region should be avoided for practical purposes because a careful control of the field and/or current is required for reliable operations. 


Note that the switched and non-switched states coexist above this second threshold, and as a result, the phase diagram shows a complex structure, as shown in Figs. \ref{fig:fig2}(c) and \ref{fig:fig2}(d). 
We notice that such a complex structure appears due to the magnetization precession around the hard ($z$) axis. 
To explain this behavior, we show the constant energy curve of the ferromagnet in the presence of the external field in Fig. \ref{fig:fig6}(a); see also Appendix \ref{sec:AppendixA}. 
There are two energetically stable states bounded by $\mathbf{m}_{0\pm}$ and $\mathbf{m}_{{\rm d}\pm}$. 
The stable states correspond to the region of $E<E_{\rm d}$ in Fig. \ref{fig:fig6}(a), 
where the saddle-point energy density is given by Eq. (\ref{eq:energy_saddle}). 
There are also two energetically unstable states, corresponding the region of $E>E_{\rm d}$. 
After turning off the current, the magnetization starts the precession on the constant energy curves, and slowly relaxes to the stable states. 
The magnetization directly relaxes to the closest stable state 
when the steady-state solution in the presence of the current locates inside one of the stable regions, 
i.e., when the energy density given by Eq. (\ref{eq:energy}) with $\mathbf{m}$ in the presence of the current is smaller than $E_{\rm d}$. 
In this case, the deterministic switching can be realized. 
On the other hand, when the steady-state solution in the presence of the current locates in one of the unstable states, 
the magnetization first starts the precession on the constant energy curves around the $z$ axis. 
In this case, even if the magnetization in the presence of the current is close to the switched state, 
the magnetization can switch back near the initial state. 
Figure \ref{fig:fig6}(b) shows an example of such a dynamics, where the red solid line represents the dynamic trajectory in the presence of the current, 
whereas the blue dotted line shows the relaxation dynamics after turning off the current. 
Figures \ref{fig:fig6}(c) and \ref{fig:fig6}(d) show the time evolutions of the magnetization components in the presence of the current and after turning off the current, respectively. 
Although the magnetization moves close to the switched state ($\mathbf{m}\simeq -\mathbf{e}_{x}$) when the current is injected, 
it turns back close to the initial state ($\mathbf{m}\simeq +\mathbf{e}_{x}$) due to the precession around the hard ($z$) axis. 
Such a dynamics is the origin of the complex structure in the phase diagram in Fig. \ref{fig:fig2}(c). 


Summarizing these considerations, the second threshold current bounding the deterministic switching to $\mathbf{m}_{0-}$ is derived 
from the condition that the steady-state solution locates inside the energetically stable region, 
and is given by (see also Appendix \ref{sec:AppendixD}) 
\begin{equation}
  j_{\rm th,\pm}
  =
  \frac{2eMd}{\hbar\vartheta}
  \frac{H_{\rm appl}H_{\rm K} \pm \sqrt{H_{\rm K}4\pi M}(H_{\rm K}+4\pi M)}{4\pi M}. 
  \label{eq:jth}
\end{equation}
We note that $j_{\rm th,+(-)}$ for a negative (positive) $H_{\rm appl}$ is the threshold current density 
to guarantee that the final state after turning off the current is $\mathbf{m}_{0-}$, 
whereas $-j_{\rm th,\pm}$ for a negative (positive) field determines the boundary that the final state is $\mathbf{m}_{0+}$. 
The black solid lines in Figs. \ref{fig:fig2}(c) and \ref{fig:fig2}(d) are Eq. (\ref{eq:jth}) and $-j_{\rm th,\pm}$ for these cases, 
indicating the validity of the formula. 

In summary, the existence of the threshold current below which the deterministic switching is achieved is found. 
We remind the readers that the critical current formula derived in Sec. \ref{sec:Critical current for instability threshold} determines the instability threshold 
to move the magnetization from the initial state. 
On the other hand, the threshold current formula derived in this section determines the relaxed state of the magnetization after turning off the current. 
Summarizing these discussions, the current density $j$ should be in the range determined by two current scales. 
For example, the current density should be in the range of $j_{\rm c,+}<j<j_{\rm th,+}$ for the deterministic switching 
when the external field is applied to the negative $z$ direction ($H_{\rm appl}<0$) and the initial state is close to the positive $x$ direction ($\mathbf{m}_{0+}$), 
as shown in Fig. \ref{fig:fig2}(c). 


\section{Comparison to other systems}
\label{sec:Comparison to other systems}

In this section, we discuss the comparison between the present and previous works. 

First, we mention that the experimentally observed value of the critical current density for the same device structure in Ref. \cite{fukami16b} is, at maximum, 
one order of magnitude smaller than that found in the present work in the small $H_{\rm appl}$ limit. 
This is because both the in-plane and perpendicular magnetic anisotropy fields are one order of magnitude smaller than those used in the present work. 
As implied from Eq. (\ref{eq:jc_zero_field}), the critical current density of the present system is roughly scaled as $\sqrt{H_{\rm K}4\pi M}$ in the small $H_{\rm appl}$ limit. 
Therefore, the critical current density in Ref. \cite{fukami16b} was small due to the smallness of $H_{\rm K}$ and $4\pi M$. 
In Ref. \cite{fukami16b}, the perpendicular demagnetization field, $4\pi M$, is suppressed by using the interfacial magnetic anisotropy effect at CoFeB/MgO interface \cite{yakata09,ikeda10,kubota12}. 


When we compare the the critical current of the present system, named as type X, with another system named type Y, 
we should first emphasize that a fair comparison is difficult because the deterministic switching in the type-X device requires the external magnetic field, 
whereas the switching in the type-Y device can be achieved without, in principle, applying the external field. 
The experimentally observed value of the critical current density for the type-Y device was on the order of $10^{7}$ A/cm${}^{2}$ \cite{liu12c}, 
which was well explained by the critical current formula derived by the macrospin assumption. 
The critical current density of the type-X system can be either small or large compared with this value, depending on the magnitude of the applied field. 
If we focus on the small field limit, however, the type Y shows a small critical current density compared with that in the type X. 
This is due to the smallness of the damping constant. 
The critical current density of the type-Y device is proportional to the damping constant $\alpha$ \cite{sun00,grollier03}, 
whereas the dependence of the critical current density of the type-X device on the damping constant is weak, as mentioned above. 
The ratio of these two critical current densities are roughly given by 
$j_{\rm c}^{\rm type\ Y}/j_{\rm c}^{\rm type\ X} \sim \alpha \sqrt{4\pi M/H_{\rm K}}$, 
where we use Eq. (\ref{eq:jc_zero_field}) and the fact that the critical current density in the type-Y device is given by  $j_{\rm c}^{\rm type\ Y}\simeq [2\alpha eMd/(\hbar\vartheta)](H_{\rm K}+2\pi M)$. 
The damping constant of the ferromagnet, such as CoFeB, conventionally used in spintronics devices is on the order of $10^{-3}-10^{-2}$ \cite{oogane06}. 
Therefore, the critical current density of type-Y device is usually smaller than that of type-X device, even though $4\pi M/H_{\rm K} \gg 1$. 
Regarding these facts, type-Y device has several advantages, such as the zero-field switching and small critical current, with respect to type-X device. 
However, we emphasize that the fast switching in type-X device is preferable for practical applications, 
compared with a slow switching in type-Y devices. 


The critical current density of the type-Z device has been found to be on the order of $10^{6}-10^{8}$ A/cm${}^{2}$ \cite{liu12b,cubukcu14,yu14,you15,torrejon15,fukami16a,oh16,brink16,lau16,kim12,kim16}, 
which are comparable to that found in type-X device. 
In both the type-X and type-Z devices, the external magnetic field is necessary for the deterministic switching. 
We note that the critical current densities of both type X and type Z are nearly independent of the damping constant \cite{lee13,taniguchi19}, 
and are dominated by the perpendicular magnetic anisotropy field. 
We should, however, note that the roles of the perpendicular magnetic anisotropy field on these devices are different. 
The perpendicular magnetic anisotropy field in type-Z device has positive sign, indicating the magnetic easy axis is parallel to the $z$ axis. 
The perpendicular magnetic anisotropy energy of type-Z device determines the thermal stability of the ferromagnet. 
On the other hand, the perpendicular magnetic anisotropy field in the type-X device, $-4\pi M$, has negative sign, indicating that the in-plane magnetized state is energetically stable. 
The thermal stability of the ferromagnet is determined by the in-plane magnetic anisotropy field $H_{\rm K}$. 
Therefore, the low critical current density and sufficient thermal stability might be simultaneously achieved in type-X device 
by suppressing the perpendicular magnetic anisotropy field by, for example, the interfacial effect \cite{yakata09,ikeda10,kubota12}. 
It should, however, be noted that the current-field range for the deterministic switching, bounded by $j_{\rm c}$ and $j_{\rm th}$ in Fig. \ref{fig:fig2}, 
is relatively narrow in the type-X device compared with that of the type-Z device; 
see Ref. \cite{taniguchi19}, where we note that the spin Hall angle used in the calculations in Ref. \cite{taniguchi19} is smaller than that in the present work. 
This result means that the type-Z device has wide tunability of the external parameters (current and field) compared with the type-X device. 


Summarizing these switching properties, the three device structures, type X, Y, and Z, have both advantages and disadvantages for the practical applications. 
An appropriate choice of the device structure, depending on the purpose, will be necessary. 


\section{Conclusion}
\label{sec:Conclusion}

In conclusion, theory of the magnetization switching caused by the spin Hall effect was developed for a ferromagnet having the easy axis parallel to the current direction. 
The analytical formula of the critical current for the magnetization switching was obtained from the theoretical condition of monotonic switching 
as a result of the competition between the spin-transfer torque and the precession torque. 
It was also found that a deterministic magnetization switching after turning off the current is achieved only when the current is in a certain range determined by another threshold current. 
Outside the current range, the final state of the magnetization becomes sensitive to the parameters such as the current and the magnetic field, 
and therefore, a precise manipulation of the magnetization becomes difficult. 
This is because the spin-transfer torque due to a large current brings the magnetization in an energetically unstable state, and induces the magnetization precession around the hard axis. 
The theoretical formula of another threshold current was also derived, showing good agreement with the result of numerical simulation. 
Summarizing these results, the current density for the deterministic switching should be in a certain range determined by two current scales. 
These results provide a solid direction for designing the three-terminal magnetic devices with in-plane easy axis. 


\section*{Acknowledgement}

The author acknowledges to Masamitsu Hayashi, Yohei Shiokawa, Shinji Isogami, Toru Oikawa, Tomoyuki Sasaki, and Seiji Mitani for valuable discussion. 
The author is also grateful to Takehiko Yorozu for his support. 
This work was supported by funding from TDK. 


\appendix


\section{Calculation details at zero temperature}
\label{sec:AppendixA}

The LLG equation was solved by the fourth-order Runge-Kutta method with the time step of $5\times 10^{-3}$ ns. 
First, we solve the LLG equation with the current (spin-transfer torque) from $t=0$ to $t=5$ $\mu$s, 
and estimate the fixed point the magnetization saturates. 
In general, the time range ($t_{\rm max}=5$ $\mu$s) is sufficient to evaluate the saturated state of the magnetization. 
After that, second, we solve the LLG equation without the current to evaluate the relaxed state of the magnetization. 
In the first step, the initial state of the magnetization is chosen to be the energetically minimum point, 
whereas the initial state of the second step is the fixed point obtained by the first step. 

The constant energy curves in Figs. \ref{fig:fig1}(b) and \ref{fig:fig6}(a) are obtained 
by solving the Landau-Lifshitz (LL) equation, where the damping constant $\alpha$ and the current density $j$ are set to be zero. 
In Fig. \ref{fig:fig1}(b), the constant energy curves, from top to bottom, are obtained by the LL equation 
with the initial conditions of $\theta=30^{\circ}$ (top red), $60^{\circ}$ (top orange), $80^{\circ}$ (top green), $83^{\circ}$ (top light blue), $85^{\circ}$ (purple), $88^{\circ}$ (red-purple), 
$97^{\circ}$ (bottom light blue), $100^{\circ}$ (bottom green), $120^{\circ}$ (bottom orange), and $150^{\circ}$ (bottom red) 
with $\varphi=0^{\circ}$ or $180^{\circ}$, 
where the zenith and azimuth angles, $\theta$ and $\varphi$, are defined as $\mathbf{m}=(\sin\theta\cos\varphi,\sin\theta\sin\varphi,\cos\theta)$. 
The corresponding energy densities estimated from Eq. (\ref{eq:energy}) are 
$10.54$, $3.42$, $0.28$, $0.06$, $-0.04$, $-0.13$ $\times 10^{6}$erg/cm${}^{3}$ for $\theta=30^{\circ}$, $60^{\circ}$, $80^{\circ}$, $83^{\circ}$, $85^{\circ}$, and $88^{\circ}$, respectively. 
Note that the energy density is the same for $\theta$ and $180^{\circ}-\theta$ in Fig. \ref{fig:fig1}(a) because the external field is absent. 
On the other hand, Fig. \ref{fig:fig6}(a) is obtained by solving the LL equation 
with the initial conditions of $\theta=30^{\circ}$ (top red), $60^{\circ}$ (top orange), $90^{\circ}$ (top green), $99^{\circ}$ (top light blue), $100^{\circ}$ (purple), $101^{\circ}$ (red-purple), 
$112^{\circ}$ (bottom light blue), $120^{\circ}$ (bottom green), and $150^{\circ}$ (bottom orange), 
corresponding to the energy densities of $17.06$, $7.17$, $-0.15$, $-0.97$, $-1.02$, $-1.06$, $-0.95$, $-0.33$, and $4.07$ $\times 10^{6}$erg/cm${}^{3}$. 
We should emphasize here that Figs. \ref{fig:fig1}(b) and \ref{fig:fig6}(a) are made to catch the overviews of the energy landscapes, 
and these values of the energy density are not explicitly used to develop the theory in the main text. 


\section{Dynamic trajectory from the initial state}
\label{sec:AppendixB}

In this Appendix, we discuss the derivations of Eqs. (\ref{eq:mx_critical_point_zero_field}) and (\ref{eq:mx_critical_point_finite_field}). 
The initial state is $\mathbf{m}_{0+}$, for convention. 

\begin{figure}
\centerline{\includegraphics[width=1.0\columnwidth]{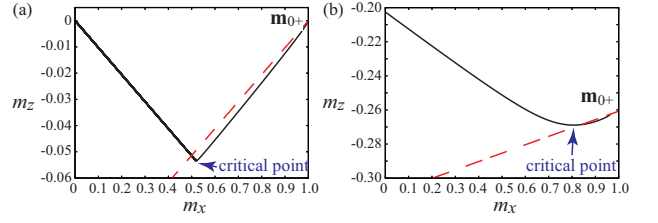}}
\caption{
         The dynamic trajectories of the magnetization in the $xz$ plane obtained from the numerical simulation of Eq. (\ref{eq:LLG}) (black solid lines) 
         and the approximated trajectories near the initial state estimated from Eqs. (\ref{eq:linear_approximation_zero_field}) and (\ref{eq:linear_approximation_finite_field}) (red dashed lines). 
         The values of the external magnetic field $H_{\rm appl}$ and the current density $j$ are 
         (a) 0 Oe and $222 \times 10^{6}$ A/cm${}^{2}$, and (b) $-5.0$ kOe and $70 \times 10^{6}$ A/cm${}^{2}$, respectively. 
         The position of the initial state is indicated as $\mathbf{m}_{0+}$, 
         whereas the location of the critical point satisfying $dm_{z}/dt=0$ is indicated by the blue arrow. 
         \vspace{-3ex}}
\label{fig:fig7}
\end{figure}

First, let us consider the zero-field case. 
We notice that the dynamic trajectory from the initial state to the critical point in the $xz$ plane is well approximated to be linear, 
as can be seen in Fig. \ref{fig:fig3}(c). 
The gradient of this approximated linear line is given by 
\begin{equation}
  \frac{\dot{m}_{z}}{\dot{m}_{x}}
  =
  \frac{H_{\rm K}m_{x}+H_{\rm s}m_{z}}{4\pi M m_{z}+H_{\rm s}m_{x}},
  \label{eq:gradient_zero_field}
\end{equation}
where the damping torque is neglected, as mentioned in the main text. 
Substituting $\mathbf{m}_{0+}=(1,0,0)$ for $H_{\rm appl}=0$ into Eq. (\ref{eq:gradient_zero_field}), 
the dynamic trajectory in the $xz$ plane with a linear approximation is described as 
\begin{equation}
  m_{z}
  =
  \frac{H_{\rm K}}{H_{\rm s}}
  \left(
    m_{x}
    -
    1
  \right).
  \label{eq:linear_approximation_zero_field}
\end{equation}
Figure \ref{fig:fig7}(a) shows the dynamic trajectory of the magnetization in the $xz$ plane obtained from the numerical simulation of Eq. (\ref{eq:LLG}) by the black solid line, 
whereas the red dashed line is Eq. (\ref{eq:linear_approximation_zero_field}). 
The values of the parameters are $H_{\rm appl}=0$ and $j=222\times 10^{6}$ A/cm${}^{2}$, as in the case of Fig. \ref{fig:fig3}(c). 
As shown, approximation obtained as Eq. (\ref{eq:linear_approximation_zero_field}) well describes the dynamic trajectory from the initial state $\mathbf{m}_{0+}$ to the critical point. 
The intersection between Eq. (\ref{eq:critical_point}) and (\ref{eq:linear_approximation_zero_field}) is the critical point at zero field, 
which is given by Eq. (\ref{eq:mx_critical_point_zero_field}). 
As mentioned in the main text, the numerically evaluated value of $m_{x}$ at the critical point satisfying $dm_{z}/dt=0$ is $0.520$, 
which is close to Eq. (\ref{eq:mx_critical_point_zero_field}). 



Next, let us extend the above discussion to the finite field case. 
We again approximate the dynamic trajectory from the initial state to the critical point to be linear. 
The gradient of the linear line is 
\begin{equation}
  \frac{\dot{m}_{z}}{\dot{m}_{x}}
  =
  \frac{H_{\rm K}m_{x}+H_{\rm s}m_{z}}{-H_{\rm appl}+4\pi M m_{z}+H_{\rm s}m_{x}}.
  \label{eq:gradient_finite_field}
\end{equation}
Therefore, the approximated dynamic trajectory in the $xz$ plane is 
\begin{equation}
  m_{z}
  =
  \frac{H_{\rm K}m_{0x}+H_{\rm s}m_{0z}}{-H_{\rm appl}+4\pi M m_{0z}+H_{\rm s}m_{0x}}
  \left(
    m_{x}
    -
    m_{0x}
  \right)
  +
  m_{0z},
  \label{eq:linear_approximation_finite_field}
\end{equation}
where $m_{0x}$ and $m_{0z}$ are given by Eq. (\ref{eq:initial_state}). 
Figure \ref{fig:fig7}(b) shows the dynamic trajectory of the magnetization in the $xz$ plane obtained from the numerical simulation of Eq. (\ref{eq:LLG}) by the black solid line, 
whereas the red dashed line is Eq. (\ref{eq:linear_approximation_finite_field}). 
The values of the parameters are $H_{\rm appl}=-5.0$ kOe and $j=70\times 10^{6}$ A/cm${}^{2}$, as in the case of Fig. \ref{fig:fig4}(b). 
Again, using Eq. (\ref{eq:linear_approximation_finite_field}) as approximation well describes the dynamic trajectory from the initial state $\mathbf{m}_{0+}$ to the critical point. 
The numerically evaluated value of $m_{x}$ at the critical point satisfying $dm_{z}/dt=0$ is $0.806$, 
which is close to the value of $0.828$ estimated as the intersection between Eq. (\ref{eq:critical_point}) and (\ref{eq:linear_approximation_finite_field}), 
as mentioned in the main text. 


\section{Calculation method at finite temperature}
\label{sec:AppendixC}

The LLG equation with the random torque given by Eq. (\ref{eq:random_torque}), is solved 
by adding the random field satisfying Eq. (\ref{eq:FDT}) to the magnetic field $\mathbf{H}$ in Eq. (\ref{eq:LLG}) \cite{taniguchi18}. 
The $k$ component ($k=x,y,z$) of the random field is given by 
\begin{equation}
  h_{k}(t)
  =
  \sqrt{
    \frac{2 \alpha k_{\rm B}T}{\gamma MV \Delta t}
  }
  \xi_{k}(t), 
\end{equation}
where the time step of the numerical simulation is $\Delta t=5 \times 10^{-3}$ ns, as mentioned in Appendix \ref{sec:AppendixA}. 
White noise $\xi$ is derived from two random numbers, $\zeta_{k}$ and $\zeta_{\ell}$, in the range of $0<\zeta_{k},\zeta_{\ell}\le 1$ 
by the Box-Muller transformation as $\xi_{k}=\sqrt{-2\log \zeta_{k}}\sin(2\pi\zeta_{\ell})$ and $\xi_{\ell}=\sqrt{-2\log \zeta_{k}}\cos(2\pi\zeta_{\ell})$. 

We calculate the LLG equation with the white noise $N=1000$ times for a given set of the applied field $H_{\rm appl}$ and the current density $j$ to evaluate the switching probability. 
Starting from the initial state in the positive $x$ region, we evaluate the $x$ component $m_{x}$ of the magnetization at $t_{\rm max}=5$ $\mu$s or $t=50$ ns. 
We note that the time step, $\Delta t$, is kept to $5\times 10^{-3}$ ns for both cases. 
We count the number $n$ of the trial where $m_{x}(t_{\rm max})<0$, and define the switching probability as $P(j)=n/N$. 
We note that the initial state is reset at each trial. 
The range of the current density for the switching probability is chosen as $j_{\rm c,+}-\Delta j_{1}\le j \le j_{\rm c,+}+\Delta j_{2}$ with the step of $\delta j=0.1 \times 10^{6}$ A/cm${}^{2}$, 
where $j_{\rm c,+}$ is the analytical value of the critical current density given by Eq. (\ref{eq:jc_finite_field}), 
whereas $\Delta j_{1}=10\times 10^{6}$ A/cm${}^{2}$ and $\Delta j_{2}=3\times 10^{6}$ A/cm${}^{2}$. 
The probability density is calculated as $dP(j)/dj \propto [P(j+\delta j)-P(j-\delta j)]/2$ to evaluate the current density where the switching probability increases most rapidly. 


\section{Boundary of deterministic switching}
\label{sec:AppendixD}

In this Appendix, we show the derivation of Eq. (\ref{eq:jth}). 
As mentioned in the main text, the deterministic switching occurs when the steady-state solution in the presence of the current locates inside one of the energetically stable regions. 
The result of the numerical simulation indicates that the steady-state solution in the presence of finite field and current locates at $m_{y}=0$; 
see, for example, Figs. \ref{fig:fig4}(a) and \ref{fig:fig4}(b). 
We notice that $dm_{x}/dt=0$ and $dm_{z}/dt=0$ are naturally satisfied at this point, as can be confirmed from the LLG equations for $m_{x}$ and $m_{z}$. 
On the other hand, using $m_{y}=0$, the condition of the steady state for $m_{y}$, $dm_{y}/dt=0$, becomes 
\begin{equation}
  -H_{\rm appl}
  m_{x}
  +
  \left(
    H_{\rm K}
    +
    4\pi M 
  \right)
  m_{x}
  m_{z}
  =
  -H_{\rm s}. 
  \label{eq:steady_state_condition_my}
\end{equation}
Although an exact solution satisfying Eq. (\ref{eq:steady_state_condition_my}) can be obtained by, for example, 
introducing a variable $\theta$ defined as $m_{x}=\sin\theta$ and $m_{z}=\cos\theta$, 
the solution is complex and is not useful. 
Instead, we use the fact that the switched state $\mathbf{m}_{0-}$ is close to the easy axis direction, $-\mathbf{e}_{x}$. 
Then, $m_{x}\simeq -1$ and $m_{z}\simeq (H_{\rm appl}+H_{\rm s})/(H_{\rm K}+4\pi M)$. 
Substituting these values into Eq. (\ref{eq:energy}), 
we find that the energy density at the steady state is approximately given by 
\begin{equation}
\begin{split}
  E_{-}
  =&
  -MH_{\rm appl}
  \left(
    \frac{H_{\rm appl}+H_{\rm s}}{H_{\rm K}+4\pi M}
  \right)
  -
  \frac{MH_{\rm K}}{2}
\\
  &
  +
  2\pi M^{2}
  \left(
    \frac{H_{\rm appl}+H_{\rm s}}{H_{\rm K}+4\pi M}
  \right)^{2}. 
\end{split}
\end{equation}
On the other hand, the saddle-point energy density is given by Eq. (\ref{eq:energy_saddle}). 
The condition that $E_{-}<E_{\rm d}$ gives Eq. (\ref{eq:jth}), which is 
the threshold current density for the deterministic switching to the stable state of $\mathbf{m}_{0-}$. 

We note that the condition to keep the magnetization near the initial state can also be derived in a similar way. 
In this case, $m_{x}\simeq +1$ and $m_{z}=(H_{\rm appl}-H_{\rm s})/(H_{\rm K}+4\pi M)$. 
The energy density at this point is 
\begin{equation}
\begin{split}
  E_{+}
  =&
  -MH_{\rm appl}
  \left(
    \frac{H_{\rm appl}-H_{\rm s}}{H_{\rm K}+4\pi M}
  \right)
  -
  \frac{MH_{\rm K}}{2}
\\
  &
  +
  2\pi M^{2}
  \left(
    \frac{H_{\rm appl}-H_{\rm s}}{H_{\rm K}+4\pi M}
  \right)^{2}. 
\end{split}
\end{equation}
The threshold current for the switching to $\mathbf{m}_{0+}$ obtained from the condition $E_{+}<E_{\rm d}$ is $-j_{\rm th,\pm}$. 


%



\end{document}